\documentclass[float,aps,prc,superscriptaddress,showpacs,twocolumn,longbibliography]{revtex4-1}

\usepackage{bm}
\usepackage{dcolumn}
\usepackage{graphicx}
\usepackage[caption=false]{subfig}
\usepackage{amsmath}
\usepackage{amssymb}
\usepackage{color}
\usepackage{epsfig}
\usepackage{hyperref}
\usepackage{MnSymbol}
\newcommand{\be}{\begin{equation}}
\newcommand{\ee}{\end{equation}}

\newcommand{\bes}{\begin{split}}
\newcommand{\ees}{\end{split}}
\newcommand{\ber}{\begin{eqnarray}}
\newcommand{\eer}{\end{eqnarray}}

\makeatletter
\def\blfootnote{\xdef\@thefnmark{}\@footnotetext}
\makeatother

\begin{document}

\title {Electric dipole polarizability from first principles calculations}

\author{M. Miorelli} \affiliation{TRIUMF, 4004 Wesbrook Mall,
  Vancouver, BC, V6T 2A3, Canada} \affiliation{Department of Physics
  and Astronomy, University of British Columbia, Vancouver, BC, V6T
  1Z4, Canada}

\author {S. Bacca} \affiliation{TRIUMF, 4004 Wesbrook Mall, Vancouver,
  BC, V6T 2A3, Canada} \affiliation{Department of Physics and
  Astronomy, University of Manitoba, Winnipeg, MB, R3T 2N2, Canada}

\author{N. Barnea} \affiliation{Racah Institute of Physics, Hebrew
  University, 91904, Jerusalem}

\author{G. Hagen} \affiliation{Physics Division, Oak Ridge National
  Laboratory, Oak Ridge, TN 37831, USA} \affiliation{Department of
  Physics and Astronomy, University of Tennessee, Knoxville, TN 37996,
  USA}

\author{G. R. Jansen} \affiliation{Physics Division, Oak Ridge National Laboratory, Oak
  Ridge, TN 37831, USA}
\affiliation{National Center for Computational Sciences, Oak Ridge National Laboratory, Oak Ridge, TN 37831 USA}

\author{G.  Orlandini} \affiliation{Dipartimento di Fisica,
  Universit\`a di Trento, Via Sommarive 14, I-38123 Trento, Italy}
\affiliation{Istituto Nazionale di Fisica Nucleare, TIFPA, Via
  Sommarive 14, I-38123 Trento, Italy}

\author{T. Papenbrock} \affiliation{Physics Division, Oak Ridge National Laboratory, Oak
  Ridge, TN 37831, USA}
\affiliation{Department of Physics and
  Astronomy, University of Tennessee, Knoxville, TN 37996, USA}

\date{\today}

\begin{abstract}
The electric dipole polarizability quantifies the low-energy behaviour
of the dipole strength and is related to critical observables
such as the radii of the proton and neutron distributions. Its
computation is challenging because most of the dipole strength lies in
the scattering continuum. In this paper we combine integral
transforms with the coupled-cluster method and compute the dipole
polarizability using bound-state techniques. Employing different
interactions from chiral effective field theory, we confirm the strong
correlation between the dipole polarizability and the charge radius,
and study its dependence on three-nucleon forces. We find good
agreement with data for the $^{4}$He, $^{40}$Ca, and $^{16}$O nuclei,
and predict the dipole polarizability for the rare nucleus $^{22}$O.
\end{abstract}

\blfootnote{This
      manuscript has been authored by UT-Battelle, LLC under Contract
      No. DE-AC05-00OR22725 with the U.S. Department of Energy. The
      United States Government retains and the publisher, by accepting
      the article for publication, acknowledges that the United States
      Government retains a non-exclusive, paid-up, irrevocable,
      world-wide license to publish or reproduce the published form of
      this manuscript, or allow others to do so, for United States
      Government purposes. The Department of Energy will provide
      public access to these results of federally sponsored research
      in accordance with the DOE Public Access Plan
      (http://energy.gov/downloads/doe-public-access-plan).}


\maketitle

\section{Introduction}
\label{sec:intro}

The electric dipole polarizability $\alpha_D$ in nuclei has been
subject of intense studies, both from the experimental and the
theoretical side. Photo-absorption studies have focused on the
determination of the giant dipole resonances (GDR) in stable nuclei,
originally interpreted as a collective motion of all protons
oscillating against all neutrons~\cite{berman1975}. The discovery of a soft peak at low
energies in neutron-rich and unstable nuclei, i.e. the pygmy dipole
resonance (PDR), has spurred a renewed interest in the electric dipole
response~\cite{kobayashi1989}. For a recent review, we refer the
reader to Ref.~\cite{aumann2013}. 

Calculations based on relativistic and non-relativistic density-functional theory pointed out that $\alpha_D$ is very strongly
correlated with the neutron-skin
thickness~\cite{Reinhard2010,Piekarewicz12,Roca-Maza2015}.  This can
be contrasted to {\it ab initio} computations based on Hamiltonians
from chiral effective field theory (EFT) that rather found a strong
correlation between the charge and the neutron radii with $\alpha_D$ in $^{48}$Ca~\cite{hagen2015}. In any
case, the dipole polarizability is sensitive to the neutron
distribution, and thereby constrains the neutron equation of state and
the physics of neutron
stars~\cite{brown2000,furnstahl2002,Tsang2012,Hebeler2014}.  The
equation of state of asymmetric nuclear matter depends on a few
parameters, such as the slope of the symmetry energy, which correlates
with GDR~\cite{Trippa2008} and PDR~\cite{Carbone2010} features.

Recent experiments measured the dipole polarizability in
$^{208}$Pb~\cite{Tamii2010}, $^{68}$Ni~\cite{Rossi2013}, and
$^{120}$Sn~\cite{krumbholz2015,Hashimoto2015}, and data for $^{48}$Ca
is presently being analyzed by the Darmstadt-Osaka collaboration.
Only scarce data exist on unstable nuclei, but recent activity was
devoted, {\it e.g.}, to $^{22,24}$O~\cite{Leistenschneider2001}.

The dipole polarizability 
\begin{equation}\label{polresp}
\alpha_D = 2\alpha \int_{\omega_{ex}}^{\infty}  d\omega~\frac{R(\omega)}{\omega}\,,
\end{equation}
where $\alpha$ is the fine structure constant, is an inverse energy weighted sum rule of the dipole response function
$R(\omega)$. Thus, the determination of the low-energy dipole strength
is crucial.  Here $\omega$ is the excitation energy and $\omega_{ex}$
is the energy of the first state excited by the dipole referred to the
ground-state.  Within one isotopic chain one expects that neutron-rich
nuclei with a significant low-lying dipole strength also exhibit a
larger polarizability than other isotopes. To both interpret recent
data and guide new experiments, it is important to theoretically map
the evolution of $\alpha_D$ as a function of neutron number. Theories
that can reliably address exotic nuclei far from the valley of
stability are needed and \textit{ab initio} methods are best positioned to
deliver both predictive power~\cite{BaccaPastore2014,Roth2014,Quaglioni16} and estimates of
the theoretical uncertainties~\cite{Carlsson16,Wesolowski15,Furnstahl15,Binder15}.

This paper is organized as follows. Section \ref{sec:th} describes the
theoretical approach based on integral transforms and the
coupled-cluster method. In Section~\ref{sec:3nf_calc} we present
results for the nuclei $^4$He, $^{16,22}$O and $^{40}$Ca. First, we
compare different computational approaches with each other.  Second,
we present results for the dipole polarizability in these nuclei based
on an interaction from chiral EFT that exhibits accurate saturation
properties~\cite{Ekstroem15}.  Third, we study correlations between
the dipole polarizability and charge radii based on a variety of
nucleon-nucleon (NN) interactions and interactions that also
include three-nucleon forces (3NFs).  Finally, we summarize our
results in Sect.~\ref{sec:conclusions}.

\section{Theoretical approach}
\label{sec:th}



The electric dipole polarizability in
Eq.~(\ref{polresp}) depends on the dipole response function
\be\begin{split}\label{resp} R(\omega) = \sumint_f &\langle \Psi_0
|\hat{\Theta}^\dag |\Psi_f\rangle\langle
\Psi_f|\hat{\Theta}|\Psi_0\rangle\delta(E_f - E_0 - \omega) .
\end{split}\ee
Here $\hat{\Theta}=\sum_{i=1}^AP_i (z_i-Z_{cm})$ is the dipole excitation operator,
where $P_i$ is the proton projection operator and $z_i$/$Z_{cm}$ the nucleon/center of mass z-coordinate, respectively.
$|\Psi_0\rangle$ is the ground state of the nucleus and
$|\Psi_f\rangle$ represents the excited states. The latter can be both
in the discrete and in the continuum region of spectrum, and this is
reflected by the combined discrete and continuum symbol
$\sumint_f$\footnote{For simplicity, in this notation we omit  the average on projections of the initial angular momentum}. From Eqs.~(\ref{polresp}) and (\ref{resp}) it is clear
that the dipole polarizability contains the information on $R(\omega)$
at all energies $\omega$, including those in the continuum. A
calculation of $\alpha_D$ would then require to be able to solve the
many-body scattering problem at such energies, which is extremely
difficult for nuclei with mass number larger than four. 

To make progress, we rewrite $\alpha_D$ as a sum rule of the response
function. Starting from Eq.~(\ref{polresp}) and using the
completeness of the Hamiltonian eigenstates
$\mathbb{I}=\sumint_{f}|\Psi_f\rangle\langle\Psi_f|$ we obtain
\be
\label{sumrule} 
\alpha_D =\langle \Psi_0|\hat{\Theta}^\dag
\frac{1}{\hat{H}-E_0} \hat{\Theta} |\Psi_0\rangle\,.  
\ee
One way to calculate $\alpha_D$ by means of the sum rule in
Eq.~(\ref{sumrule}) is to represent the Hamiltonian on a finite basis of
$N$ basis functions $|n\rangle$.  After diagonalization of the
Hamiltonian matrix $H_{n,n'}$, one obtains its $N$ eigenstates
$|\beta\rangle$ and eigenvalues $E_\beta$, and Eq.~(\ref{sumrule})
becomes 
\be
\label{sumrule_comp} \alpha_D =\sum_{\beta}^{N} \langle
\Psi_0|\hat{\Theta}^\dag | \beta \rangle \langle \beta
|\frac{1}{E_\beta-E_0} |\beta \rangle \langle \beta | \hat{\Theta}
|\Psi_0\rangle\,.  
\ee 
Increasing $N$ yields an increasingly more accurate representation of
the eigenfunctions $|\beta\rangle$ and eigenvalues $E_\beta$ of
$\hat{H}$, and eventually the value of $\alpha_D$ would converge. In
practical cases, however, the truncated basis states $|n\rangle$ used
to represent the Hamiltonian are discrete and have a finite
norm. Because the spectrum of $\hat{H}$ has both a
discrete and a continuum part, one may question the use of such a
discrete basis. Similarly to Ref.~\cite{LSR}, we will show that this approach is rigorous and works quite well also within coupled-cluster theory.

\subsection{Integral transforms}
\label{subsec:itf}
Integral transforms reduce the continuum problem of calculating
$R(\omega)$ to the solution of a bound-state-like
problem~\cite{Efros85,Efros94,Efl07}.  In such an approach, one first
calculates the integral transform ${\mathcal I}(\sigma)$ of the
response function. In a second step, one might invert the integral
transform to obtain the response function $R(\omega)$, or one might
compute relevant observables (such as the dipole polarizability)
directly from the integral transform. Here, we will use the Stieltjes
integral transform \cite{Efros93} for the direct computation of the
dipole polarizability.

\begin{figure}[!tb]
\includegraphics[width=\linewidth,clip=]{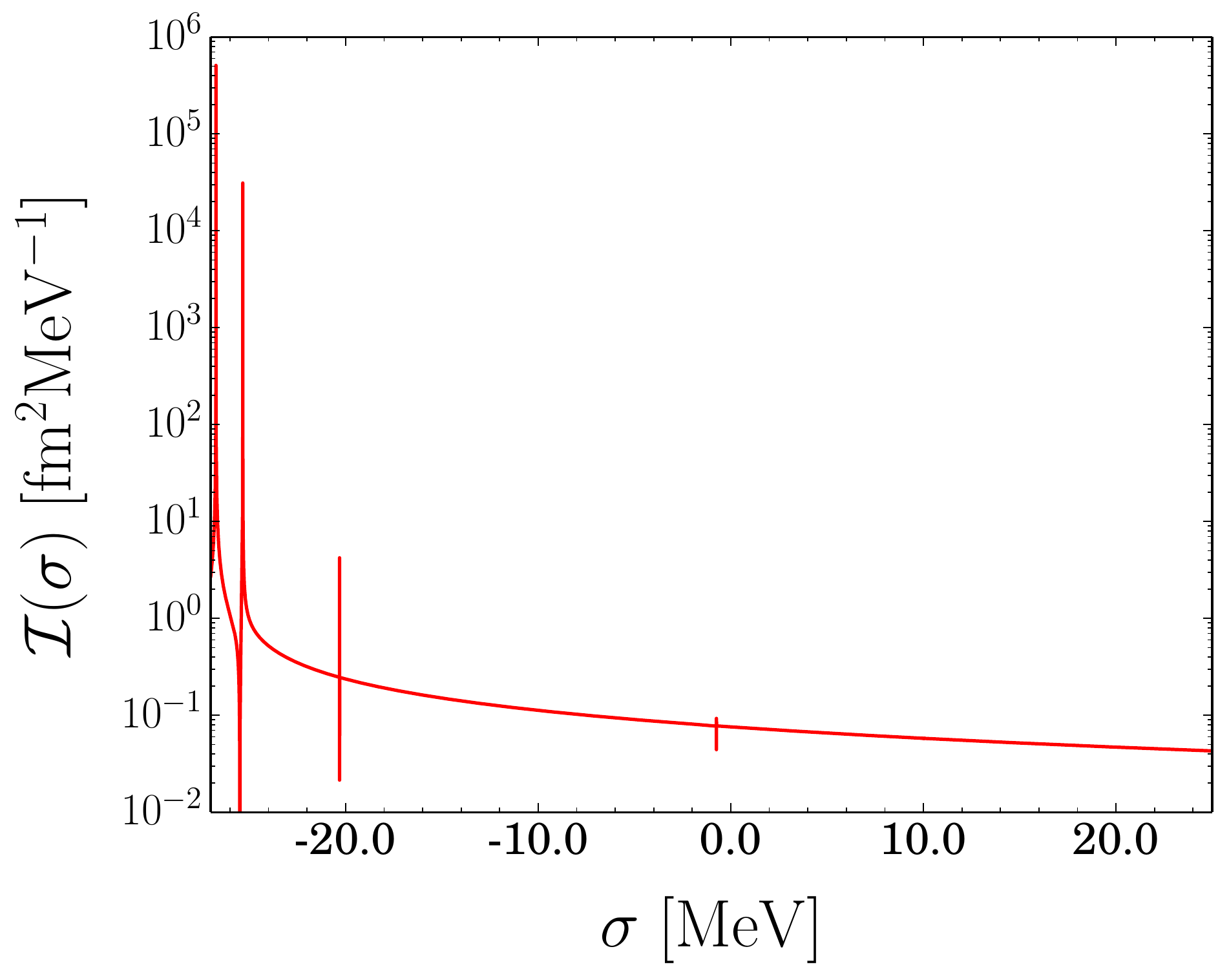}
\caption{(Color online) The Stieltjes integral transform
  $\mathcal{I}(\sigma)$ as a function of $\sigma$ in the case of
  $^{4}$He.}
\label{fig:stieltjes}
\end{figure}

The Stieltjes integral transform reads 
\be\label{stieltjes}
\mathcal{I}(\sigma) =\int\frac{R(\omega)}{\omega+\sigma}d\omega, 
\ee
with $\sigma$ real and positive. Using the completeness on the
Hamiltonian eigenstates and the definition of the response function from
Eq.~(\ref{resp}) yields
\be
\label{stieltjes2}\begin{split}
  \mathcal{I}(\sigma) &=\langle \Psi_0|\hat{\Theta}^\dag
  \frac{1}{\hat{H}-E_0+\sigma} \hat{\Theta} |\Psi_0\rangle\\ &=
  \langle\Psi_0|\hat{\Theta}^\dag|\tilde{\Psi}(\sigma)\rangle,
\end{split}
\ee
where we have defined 
\be\label{psitilde}
|\tilde \Psi(\sigma)\rangle\equiv \frac{1}{\hat{H}-E_0+\sigma} \hat{\Theta} |\Psi_0\rangle.
\ee
The function $|\tilde \Psi(\sigma)\rangle$ is the solution of the
following Schr\"odinger-like equation with a source
\be
\label{Schr_like_eq} (\hat{H}-E_0+\sigma)|\tilde
\Psi(\sigma)\rangle = \hat{\Theta} |\Psi_0\rangle.  
\ee 
Since $\sigma >0$, and for large inter particle distances
$|\Psi_0\rangle \rightarrow 0$, one has that asymptotically -- and for
non singular operators $\hat{\Theta}$ -- $|\tilde \Psi(\sigma)\rangle$
should satisfy a Schr\"odinger equation with eigenvalues smaller than
$E_0$. This implies that $|\tilde \Psi(\sigma)\rangle \rightarrow 0$
asymptotically, namely it has bound state-like asymptotic
conditions. We are therefore allowed to calculate
$\mathcal{I}(\sigma)$ using a bound-state basis expansion, \textit{i.e.} an $L_2$ square integrable basis such as harmonic oscillator functions. Noticing
that Eq.~(\ref{stieltjes2}) differs from Eq.~(\ref{sumrule}) only by
the presence of $\sigma >0$, we proceed as it was described above,
namely using a representation on a bound state basis and increasing
the number $N$ of basis functions up to convergence.  Then the value
of $\alpha_D$ can be obtained as
\be
\label{polstil}
\alpha_D=2\alpha\lim_{\sigma\to 0^+}\mathcal{I}(\sigma)\,,
\ee
avoiding the continuum problem. The limit taken with positive $\sigma$
is crucial not only to allow the use of a bound state basis, but also
because it avoids poles (we recall that $E_0$ is negative).  
For $\sigma<0$ poles will certainly be present, presumably at different places depending on the basis. We indeed observe several poles  in the region of $\sigma<0$, while the curve is persistently smooth for $\sigma\geq 0$ in Fig.~\ref{fig:stieltjes}, where we show $\mathcal{I}(\sigma)$ for $^4$He calculated with a realistic interaction~\cite{Ekstroem15}, as detailed later.  We choose $^4$He, where calculations are faster and can be benchmarked with few-body methods.

Below we will use an implementation of Eq.~(\ref{polstil}) to compute
the dipole polarizability of heavier nuclei. To test this approach, we will also compare it to $\alpha_D$ obtained by the dipole response function as in Eq.~(\ref{polresp}). If one were able to invert the Stieltjes transform, one could obtain $R(\omega)$ to calculate the 
integral in Eq.~(\ref{polresp}). Unfortunately, the inversion of this integral transform presents the typical difficulties of an ill-posed problem. In fact in Ref.~\cite{Efros93} it was shown that inversions performed with the regularization method~\cite{Tikhonov} generate 
quite different responses, all compatible with the same Stieltjes transform within numerical errors.
Therefore, we will employ the much more suitable Lorentz integral transform (LIT)~\cite{Efros94,Efl07} 

\be
\label{litdef}
L(\sigma,\Gamma) = \frac{\Gamma}{\pi}\int\frac{R(\omega)}{(\omega -
  \sigma)^2 + \Gamma^2}d\omega, 
\ee 
where $\Gamma,\sigma \in \mathbb{R}$ and $\Gamma >
0$. The Lorentzian kernel $L(\sigma,\Gamma)$ is peaked at $\sigma$ and
has the width $\Gamma$. The LIT can be much more easily inverted to yield the response function,  because the width $\Gamma$ introduces a finite resolution.
Thus, the response function is smeared only in a narrow region of space determined by the width $\Gamma$.

The calculation the Lorentz transform proceeds as for the Stieltjes transform, using the definition of response function and the
completeness of the eigenfunctions of the Hamiltonian. One finds

\be
\label{lit}
L(z) = \frac{\Gamma}{\pi}\langle\Psi_0|\hat{\Theta}^\dag\frac{1}{(\hat{H} -z^*)}\frac{1}{(\hat{H} -z)}\hat{\Theta}|\Psi_0\rangle,
\ee
with $z = E_0 + \sigma +i\Gamma$. The LIT can be rewritten in a form
that resembles Eq.~(\ref{stieltjes2}) as
\be
\label{lit2}\begin{split}
L(z) &= \frac{1}{\pi}\mathfrak{Im}\left[\langle\Psi_0|\hat{\Theta}^\dag\frac{1}{(\hat{H} - z)}\hat{\Theta}|\Psi_0\rangle\right]\\
&=\frac{1}{\pi}\mathfrak{Im}\left[\langle\Psi_0|\hat{\Theta}^\dag|\tilde\Psi(z)\rangle\right]\,.
\end{split}
\ee
Here we defined the function
\be
\label{shrodlit}
|\tilde\Psi(z)\rangle \equiv  \frac{1}{\hat{H}- z} \hat{\Theta}|\Psi_0\rangle.
\ee
Similarly as for Eq.~(\ref{psitilde}), $|\tilde{\Psi}(z)\rangle$ has a
bound-state-like nature and a finite norm
\be\label{norm}
\langle\tilde{\Psi}(z)|\tilde{\Psi}(z)\rangle=L(z)=
\frac{\Gamma}{\pi}\int\frac{R(\omega)}{(\omega - \sigma)^2 + \Gamma^2}d\omega <\infty.
\ee
A couple of remarks are in order here. First, we note that the positive
parameter $\sigma$ enters in the Stieltjes and Lorentz transforms with
a minus and a plus sign respectively. While in
the Stieltjes transform the bound-state-like nature of
$|\tilde{\Psi}\rangle$ is due to that minus sign, in the Lorentz case
it is due to the presence of the imaginary part $\Gamma$.  Second,
in the limit $\Gamma \rightarrow 0$ 
the Lorentzian kernel becomes a delta function
\be\label{litdelta}
L(\sigma,\Gamma\to 0) = \int R(\omega)\delta(\omega - \sigma)d\omega=R(\sigma).
\ee
This allows us to estimate the dipole polarizability also using
Eq.~(\ref{litdelta}) together with Eq.~(\ref{polresp})
\be\label{poldelta}
\alpha_D = 2\alpha\int \frac{L(\sigma,\Gamma\to 0)}{\sigma}d\sigma.
\ee 
However, in L($\sigma$,$\Gamma$) one must be careful in taking smaller and smaller $\Gamma$ since the convergence in the model space expansion becomes increasingly difficult.

\subsection{Coupled-cluster implementation}
\label{subsec:lanzos}

In this Subsection we will compute the dipole polarizability via
Eq.~(\ref{polstil}) with the coupled-cluster method. This
calculation proceeds similarly as done for the LIT in
Refs.~\cite{Bacca13,LITCC}.

Coupled-cluster
theory~\cite{coester1958,coester1960,kuemmel1978,mihaila2000b,dean2004,wloch2005,hagen2010b,binder2013b}
is based on the exponential ansatz for the ground state
\be\label{expansatz}
|\Psi_0\rangle = e^{\hat{T}}|0_R\rangle, 
\ee
see Refs.~\cite{bartlett2007,hagen2014} for recent reviews. Here,
$|0_R\rangle$ is a reference product state, and the cluster operator
$T$ introduces particle-hole (p-h) excitations into the
reference. Using second quantization, and normal ordering the dipole
excitation operator with respect to the reference state yields the
response function~\cite{LITCC}
\be\label{noresp} R(\omega) = \sum_n \langle
0_L|\overline{\Theta}^\dag|n_R\rangle\langle
n_L|\overline{\Theta}|0_R\rangle\delta(\Delta E_n - \Delta E_0 -
\omega).
\ee 
Here $\Delta E_n$, $\Delta E_0$ are the correlation
energies of the $n$th-excited state and ground-state respectively,
and solve 
\be
\begin{split}\label{CCenergy}
\overline{H}|0_R\rangle = \Delta
E_0|0_R\rangle\ \ \ \ &\text{or}\ \ \ \ \langle 0_L|\overline{H} =
\langle 0_L|\Delta E_0,\\ \overline{H}|n_R\rangle = \Delta
E_n|n_R\rangle\ \ \ \ &\text{or}\ \ \ \ \langle n_L|\overline{H} =
\langle n_L|\Delta E_n.
\end{split}
\ee
Here we used similarity-transformed operators via
\begin{equation}\label{simtransf}
\overline{O} = e^{-\hat{T}}\hat{O}_Ne^{+\hat{T}},
\end{equation}
and $\hat{O}_N$ is the normal-ordered form of any
operator $\hat{O}$, \textit{e.g.} $\hat{H}$ or $\hat{\Theta}$. Substituting Eq.~(\ref{noresp}) in Eq.~(\ref{stieltjes}),
and making use of the expressions in Eq.~(\ref{CCenergy}) yields
\be
\label{CCstieltjes} \mathcal{I}(\sigma) = \langle
0_L|\overline{\Theta}^\dag\frac{1}{\overline{H} - \Delta E_0 +
  \sigma}\overline{\Theta}|0_R\rangle.  
\ee
This equation resembles Eq.~(\ref{stieltjes2}), when
operators are replaced by their similarity transformed
counterparts, and one needs to distinguish between left and right states because
of the non-Hermitian nature of the excitation operator $\hat{T}$. We 
proceed as in Subsection~\ref{subsec:itf}, and define a state
$|\tilde{\Psi}(\sigma)\rangle$ as the solution of 
\be\label{LITCC_eq}
(\overline{H} - \Delta E_0 + \sigma)|\tilde{\Psi}_R(\sigma)\rangle =
\overline{\Theta}|0_R\rangle.  
\ee
Eq.~(\ref{LITCC_eq}) resembles Eq.~(15) in \cite{LITCC} and can be
solved using the equation-of-motion coupled-cluster method for excited
states \cite{Stanton93}. In this approach, one regards
\be\begin{split}\label{cceom1}
|\tilde{\Psi}_R(\sigma)\rangle =
&\hat{\mathcal{R}}(\sigma)|0_R\rangle\equiv \left(r_0(\sigma) +
\sum_{i,a}r^a_i(\sigma) \hat{c}^\dag_a \hat{c}_i +\right.\\ &\left. +
\frac{1}{4}\sum_{i,j,a,b}r^{ab}_{ij}(\sigma)\hat{c}_a^\dag\hat{c}^\dag_b\hat{c}_j\hat{c}_i
+ ...\right)|0_R\rangle\\ \equiv &\sum_\alpha\hat{C}_\alpha
r_\alpha(\sigma)|0_R\rangle\equiv
\hat{\mathbf{C}}\cdot\mathbf{r}(\sigma)|0_R\rangle,
\end{split}
\ee
as an excited state of the similarity-transformed Hamiltonian $\overline{H}$ based on
p-h excitations of the reference. In the last line of Eq.~(\ref{cceom1}) 
the index $\alpha$ labels the $0$p-$0$h,
$1$p-$1$h, $2$p-$2$h, ... states 
\be\label{states}
|\Phi_\alpha\rangle \equiv |0_R\rangle, |\Phi_i^a\rangle, |\Phi^{ab}_{ij}\rangle, \ldots . 
\ee 
We also defined the column vector $\mathbf{r}(\sigma)$ with elements 
$r_0(\sigma),r^a_i(\sigma), r^{ab}_{ij}(\sigma), ...$ and a row
vector $\hat{\mathbf{C}}$ whose elements are strings of normal-ordered
creation and annihilation operators. Combining Eq.~(\ref{LITCC_eq}) with
Eq.~(\ref{CCenergy}) and the linear ansatz for
$|\tilde{\Psi}(\sigma)\rangle$, the Stieltjes transform becomes
\be\label{matrix_eq}
\mathcal{I}(\sigma) = \langle 0_L|\overline{\Theta}^\dag\hat{\mathcal{R}}(\sigma)|0_R\rangle = \mathbf{S}^L\mathbf{M}(\sigma)^{-1}\mathbf{S}^R .
\ee
Here $\mathbf{S}^L$ and $\mathbf{S}^R$ are row- and column-vectors
respectively with elements 
\be
\begin{split}
&S^R_\alpha = \langle\Phi_\alpha|\overline{\Theta}|0_R\rangle,\\
&S^L_\alpha = \langle0_L|\overline{\Theta}^\dag|\Phi_\alpha\rangle,
\end{split}
\ee
and $\mathbf{M}$ is a matrix with elements
\be
M_{\alpha\beta}(\sigma) = \langle \Phi_{\alpha} |\left[\overline{H},\hat{C}_\beta\right]|0_R\rangle + \sigma\delta_{\alpha\beta}.
\ee
The right-hand side of Eq.~(\ref{matrix_eq}) can be calculated using the Lanczos
procedure. Because we are dealing with non-Hermitian operators, we
have to make use of the generalized Lanczos algorithm for
non-symmetric matrices \cite{Cullum98}. In this approach, one first
defines two pivot vectors
\be\label{vecdef}\begin{split}
&\mathbf{v}_0 = \frac{\mathbf{S}^R}{\sqrt{\mathbf{S}^L\cdot\mathbf{S}^R}},\\
&\mathbf{w}_0 = \frac{\mathbf{S}^L}{\sqrt{\mathbf{S}^L\cdot\mathbf{S}^R}},
\end{split}
\ee
and repeated application of the matrix $\mathbf{M}(\sigma)$ generates the Lanczos basis
in which $\mathbf{M}$ is tri-diagonal
\be\label{trimat}
\mathbf{M}(\sigma)=
\begin{pmatrix}
a_0 - \sigma& b_0 & 0 & 0 & \cdots\\
b_0 & a_1 -\sigma & b_1 & 0 & \cdots\\
0 & b_1 & a_2 - \sigma & b_2 & \cdots\\
0 & 0 & b_2 & a_3 - \sigma & \cdots\\
\vdots & \vdots & \vdots & \vdots & \ddots
\end{pmatrix}.
\ee
In what follows, we employ the matrix $\mathbf{M}$ in the Lanczos basis. 
We note that $\mathbf{S}^L\cdot\mathbf{S}^R = \langle
0_L|\overline{\Theta}^\dag\overline{\Theta}|0_R\rangle$ and find the expression 
\be\label{stilcontfrac}
\mathcal{I}(\sigma) = \langle 0_L|\overline{\Theta}^\dag\overline{\Theta}|0_R\rangle x_{00}(\sigma),
\ee
for the Stieltjes integral transform. Here
\be\label{contfrac}
x_{00}(\sigma) = \mathbf{w}_0\left[\mathbf{M}(\sigma)\right]^{-1}\mathbf{v}_0.
\ee
From the identity $\mathbb{I} =
\mathbf{M}(\sigma)[\mathbf{M}(\sigma)]^{-1}$, one finds the linear
system
\be
\sum_\beta {M}_{\alpha\beta}(\sigma)x_{\beta 0}(\sigma) = \delta_{\alpha 0},
\ee
where we defined $x_{\beta 0}(\sigma) =
[\mathbf{M}(\sigma)^{-1}]_{\beta 0}$. Using Cramer's rule to solve the
linear system, we find that $x_{00}(\sigma)$ is given by the continued
fraction
\be\label{contfrac2}
x_{00}(\sigma) = \frac{1}{(a_0 - \sigma) - \frac{b_0^2}{(a_1 - \sigma) - \frac{b_1^2}{(a_2 - \sigma) - \cdots}}},
\ee
and finally Eq.~(\ref{stilcontfrac}) becomes
\be\label{stilcontfrac2}
\mathcal{I}(\sigma) = \langle 0_L|\overline{\Theta}^\dag\overline{\Theta}|0_R\rangle\left\{\frac{1}{(a_0 - \sigma) - \frac{b_0^2}{(a_1 - \sigma) - \frac{b_1^2}{(a_2 - \sigma) - \cdots}}}\right\}.
\ee
Then, from Eq.~(\ref{polstil}), one finds that the electric dipole
polarizability is the  continued fraction
\be\label{polstilCC}
\alpha_D =2\alpha\langle 0_L|\overline{\Theta}^\dag\overline{\Theta}|0_R\rangle\lim_{\sigma\to 0^+}\left\{\frac{1}{(a_0 + \sigma) - \frac{b_0^2}{(a_1 + \sigma) - \frac{b_1^2}{(a_2 + \sigma) - \cdots}}}\right\} ,
\ee
which is equivalent to the Lanczos sum rule of Ref.~\cite{LSR}. We note that Eq.~(\ref{polstilCC}) is an exact result if the operators $\hat{T}$ and $\hat{\mathcal{R}}$ are expanded up to $A$p-$A$h excitations in a nucleus with mass number $A$.  However, in practical calculations $\hat{T}$ and $\hat{\mathcal{R}}$ are truncated since a full expansion is not feasible due to the very high computational cost. In this paper we truncate $\hat{T}$ and $\hat{\mathcal{R}}$ at singles-and-doubles excitations. However, we remind the reader that such a truncation includes exponentiated $1$p-$1$h and $2$p-$2$h excitations. The exponent yields also products of higher order.  As the GDR consists of a superposition of $1$p-$1$h excitations, a truncation at singles-and-doubles only is expected to be a good approximation. Similarly, the dipole polarizability is most sensitive to the GDR. 

Summarizing, we presented three different methods to evaluate the electric dipole polarizability: (i) compute the LIT for the dipole response, obtain $R(\omega)$ from its inversion -- with inversions  performed as described in Ref.~\cite{Bacca13,Andreasi05,Efros99} -- and compute the dipole  polarizability from Eq.~(\ref{polresp}); (ii) use Eq.~(\ref{poldelta}) for $\Gamma\to 0$.  (iii) use the continued fraction as in Eq.~(\ref{polstilCC}).  The second method is in principle a discretization of the continuum and it will be interesting to compare it with the other two methods.

\section{Results}
\label{sec:3nf_calc} 

In Ref.~\cite{Bacca13,LITCC} coupled-cluster results for the dipole
response in $^{4}$He were benchmarked against virtually exact results
from the effective interaction hyperspherical harmonics~\cite{EIHH,Goerke12} method. 
Those calculations were based on NN forces~\cite{Entem03} from chiral EFT. In this paper we augment the Hamiltonians to include 3NFs from chiral EFT.

First, we check the convergence of our results with respect to model-space parameters and compare the three different calculational approaches for the dipole polarizability using the NNLO$_{\rm sat}$ interaction~\cite{Ekstroem15}.
Then, we compare to experimental data for $^4$He, $^{16}$O and $^{22}$O. Finally, we explore correlations of the dipole polarizability with the charge radius in $^{16}$O and $^{40}$Ca by employing a variety of Hamiltonians. In addition to NNLO$_{\rm{sat}}$ and the family of interactions from Ref.~\cite{Hebeler11}, we also use a large set of realistic NN potentials~\cite{CDBONN,Entem03,bogner2007,bogner2003,AV18} to probe systematic uncertainties in the underlying Hamiltonians.

When adding 3NFs, we use a Hartree-Fock basis built on 15 major harmonic oscillator shells. We vary the model space size up to $N_{\rm max}=14$ and we truncate the 3NFs matrix elements at $E_{\rm 3max}=N_{\rm max}$ for $^4$He and $^{16,22}$O. For our purposes, this truncation provides well-enough converged results. In fact, for the more challenging neutron-rich $^{22}$O nucleus, increasing $E_{\rm 3max}$ to 16 leads to a variation in energy of only 400~keV. For the calculations in $^{40}$Ca with the NNLO$_{\rm sat}$ interaction and the Hamiltonians from Ref.~\cite{Hebeler11} we employed the same $N_{\rm max}$ and $E_{\rm 3max}$ truncations used for $^{48}$Ca by ~\citeauthor{hagen2015} in Ref.~\cite{hagen2015}.

\subsection{The $^4$He nucleus}
\label{subsec:4He} 

\begin{figure}[!tb]
\includegraphics[width=\linewidth,clip=]{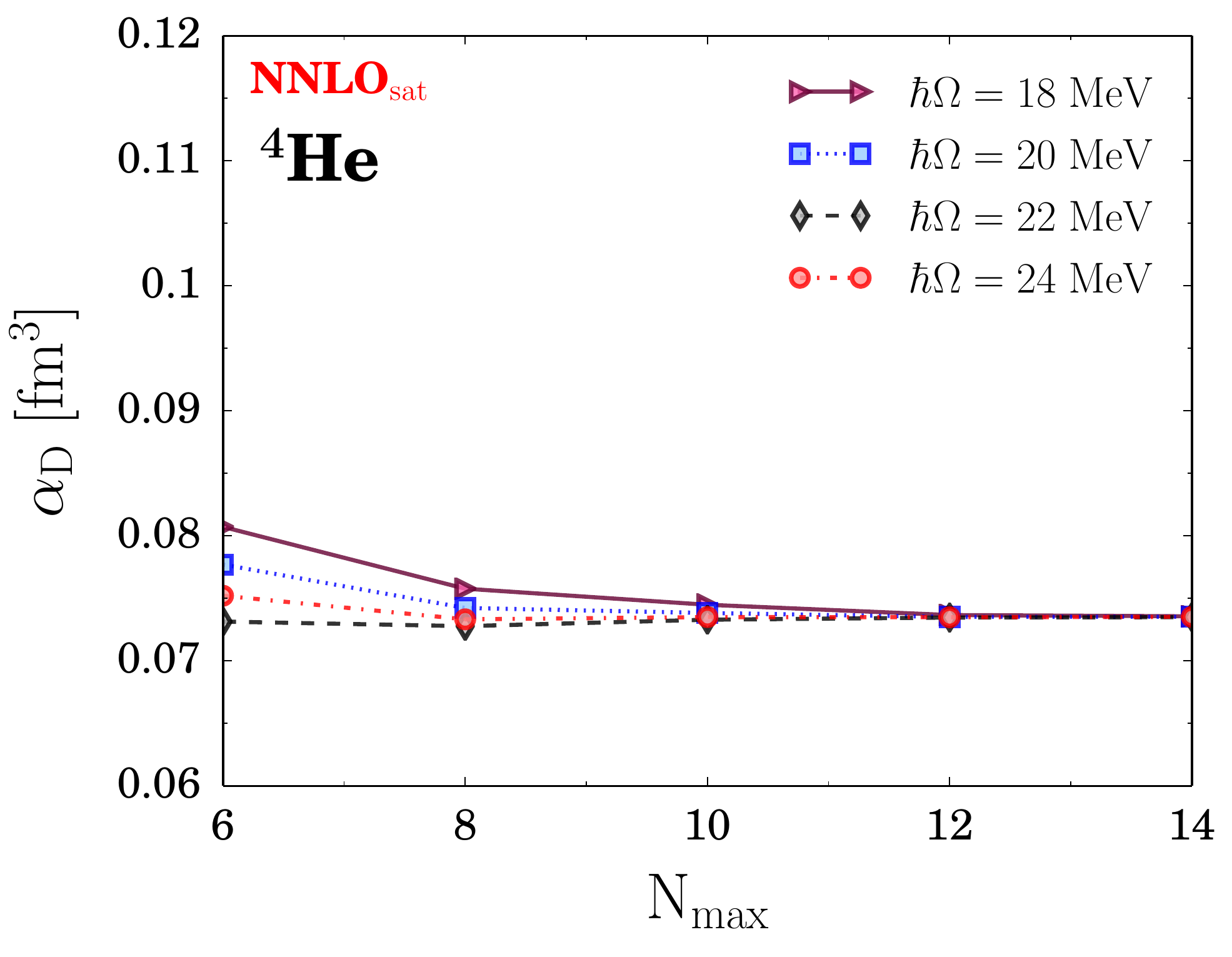}
\caption{(Color online) The electric dipole polarizability in $^{4}$He
  as a function of the model space size $N_{\rm max}$. Curves for
  different values of $\hbar\Omega$, the underlying harmonic
  oscillator frequency, are shown.}
\label{fig:fig_pol_conv_4He}
\end{figure}

Figure~\ref{fig:fig_pol_conv_4He} shows the electric dipole
polarizability of $^4$He obtained from the continued fraction of
Eq.(\ref{polstilCC}) with the NNLO$_{\rm{sat}}$ interaction, as a
function of the model space size $N_{\rm max}$. The four curves represent
calculations with different values of oscillator frequency
$\hbar\Omega$.  The convergence in $N_{\rm max}$ is excellent and
independence on $\hbar \Omega$ is reached with $N_{\rm max}=14$. The uncertainty 
at $N_{\rm max}=14$ for the different values of $\hbar\Omega$ is about $0.1\%$.

\begin{figure}[!tb]
\includegraphics[width=\linewidth,clip=]{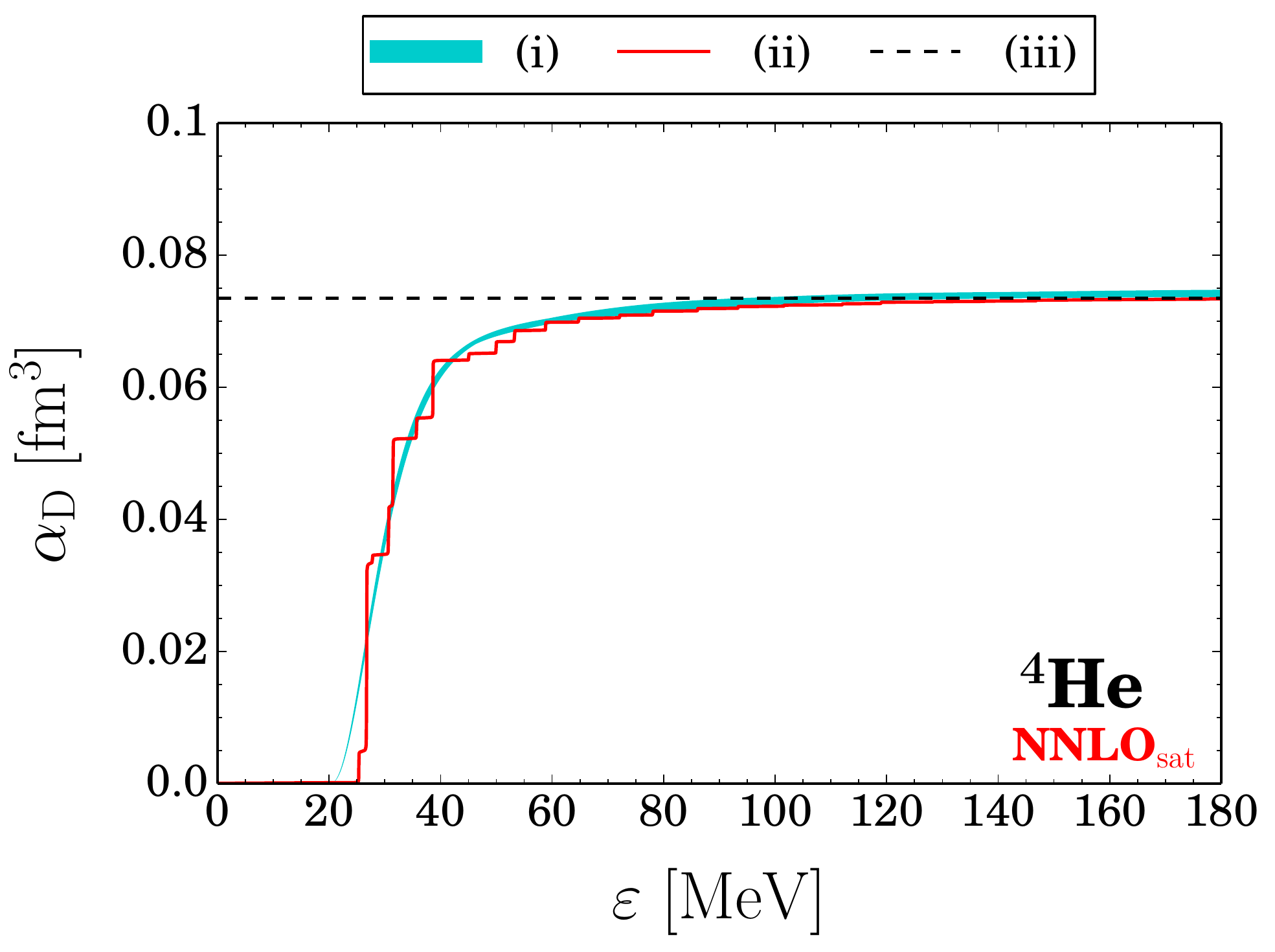}
\caption{(Color online) The electric dipole polarizability
  $\alpha_D(\varepsilon)$ in $^{4}$He as a function of the integration
  energy $\varepsilon$: (i) using the LIT and Eq.~(\ref{polresp}) in
  blue (band); (ii) using Eq.~(\ref{poldelta}) in red (solid); (iii)
  using the continued fraction of Eq.~(\ref{polstilCC}) in black
  (dashed). Calculations are performed for $\hbar \Omega=22$ MeV and
  $N_{\rm max}=14$.}
\label{fig:fig_pol_vs_e_4He}
\end{figure}

Let us compare the three different ways to calculate the dipole
polarizability for $^4$He as described at the end of
Section~\ref{subsec:itf}. Equations~(\ref{polresp}) and
(\ref{poldelta}) require an integration in energy and we present
$\alpha_D(\varepsilon)$ where $\varepsilon$ is the upper limit of the
integration.  Figure~\ref{fig:fig_pol_vs_e_4He} shows the results.
The blue band shows method (i), i.e.  $\alpha_D$ is obtained from
integrating Eq.~(\ref{polresp}), and $R(\omega)$ stems from an
inversion of the LIT.  Here, the width of the blue band is an estimate
of the uncertainty involved in the inversion procedure. The red solid
line shows method (ii), i.e.  $\alpha_D$ obtained from the LIT at
small $\Gamma$ using Eq.~(\ref{poldelta}). The black dashed line shows
the method (iii), i.e. $\alpha_D$ obtained using the continued
fraction in Eq.~(\ref{polstilCC}).

We note that the different methods yield the same dipole
polarizability.  The integration methods (i) and (ii) exhibit a
similar dependence on the integration range, the difference being that
the former is smooth while the latter increases in steps. Here, method
(ii) has the advantage of a sharper defined threshold. We also note
that the dependence on the integration range is useful for comparisons
with data for experiments that probe only a limited region of the
energy spectrum.

\subsection{The $^{16}$O nucleus}
\label{sec:16O} 

\begin{figure}[!ht]
\includegraphics[width=\linewidth,clip=]{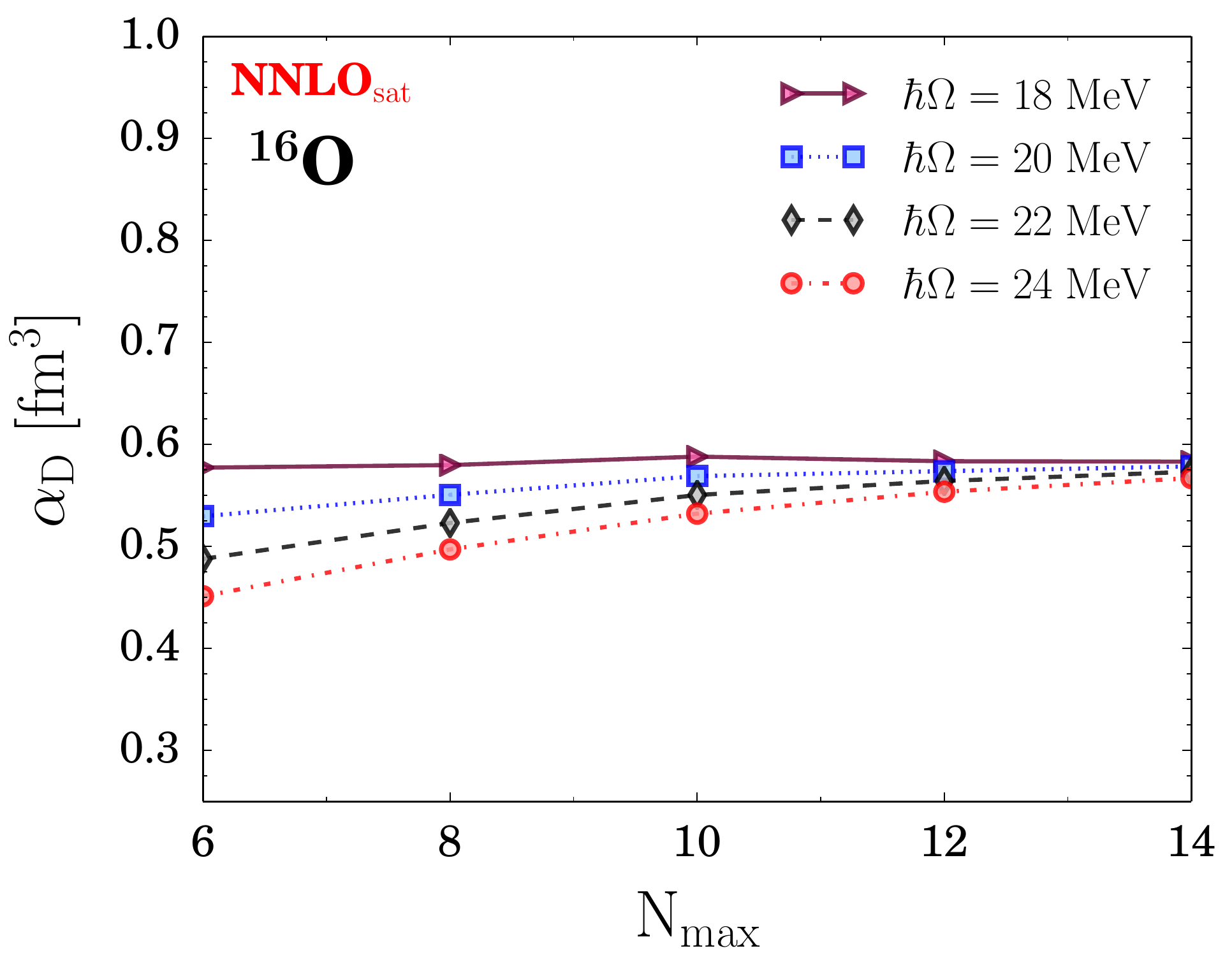}
\caption{(Color online) Electric dipole polarizability in $^{16}$O as
  a function of the model space size $N_{\rm max}$ for different values of
  $\hbar\Omega$.}
\label{fig:fig_pol_conv_16O}
\end{figure}

\begin{figure}[!ht]
\includegraphics[width=\linewidth,clip=]{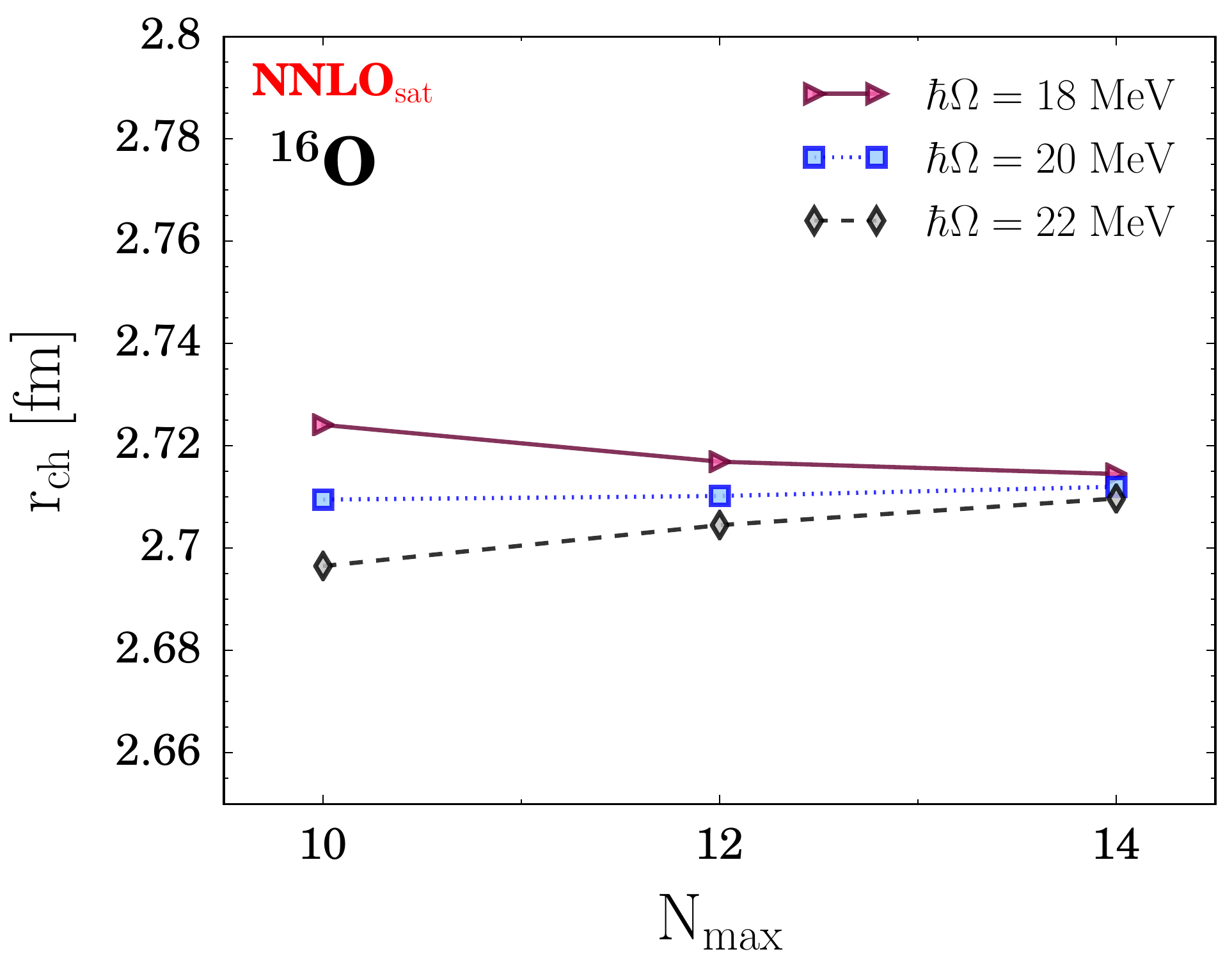}
\caption{(Color online) Charge radius in $^{16}$O as
  a function of the model space size $N_{\rm max}$ for different values of
  $\hbar\Omega$.}
\label{fig:fig_rch_conv_16O}
\end{figure}

Figure~\ref{fig:fig_pol_conv_16O} shows the electric dipole
polarizability in $^{16}$O as a function of the model space size calculated with
the NNLO$_{\rm{sat}}$ interaction, while Figure~\ref{fig:fig_rch_conv_16O} shows the same for the charge radius, which has been obtained from the point-proton radius taking into account contributions from nucleonic charge radii (see Ref.~\cite{hagen2015} for details). We observe that the curves for different $\hbar\Omega$ values converge very nicely and only a small residual $\hbar \Omega$-dependence remains at the largest model space size $N_{\rm max}=14$. Based on the spread of the different $\hbar\Omega$ curves for $N_{\rm max}=14$, we obtain a conservative error of $2.8\%$ for the electric dipole polarizability and a conservative error of $0.7\%$ for the charge radius.

\begin{figure}[!ht]
\includegraphics[width=\linewidth,clip=]{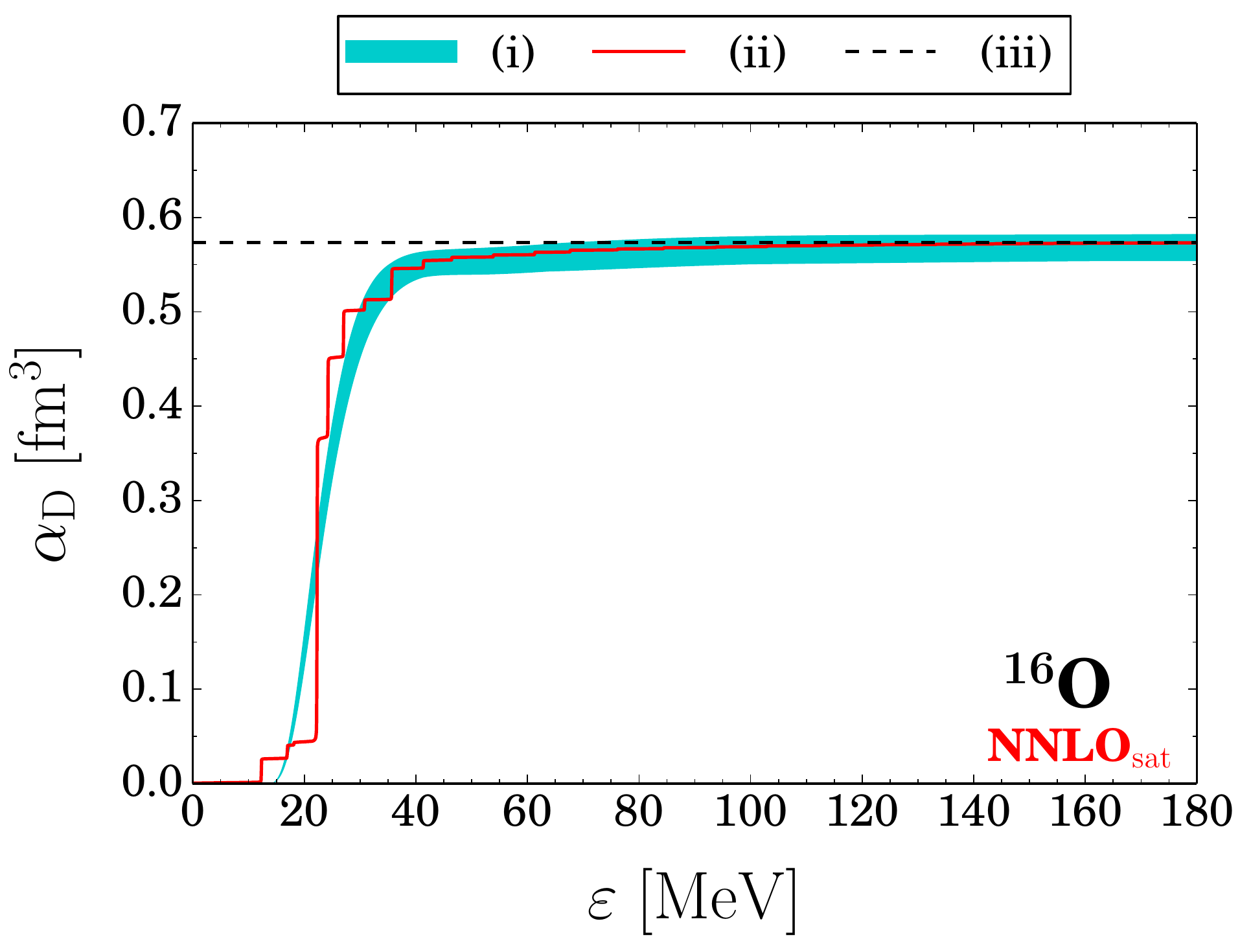}
\caption{(Color online) The electric dipole polarizability
  $\alpha_D(\varepsilon)$ in $^{16}$O as a function of the integration
  limit $\varepsilon$.  The blue band (i) is obtained integrating the
  weighted response function as in Eq.~(\ref{polresp}); the red solid
  curve (ii) is calculated integrating the weighted LIT at small
  $\Gamma$ as in Eq.~(\ref{poldelta}); the black dashed line (iii) is
  obtained from the continued fraction of
  Eq.~(\ref{polstilCC}). Calculations are performed with $N_{\rm
    max}=14$ and $\hbar \Omega=22$ MeV.}
\label{fig:fig_polVSen_n2losat_16O}
\end{figure}

Figure~\ref{fig:fig_polVSen_n2losat_16O} compares the results from the
three methods to obtain the polarizability for $^{16}$O. The blue band (i) shows the integration as
in Eq.~(\ref{polresp}) of the weighted response function, and the
width of the band takes into account the uncertainty of the inversion.
The red solid line (ii) refers to the integration of the weighted LIT
with Eq.~(\ref{poldelta}).  The black dashed line (iii) is the
reference value calculated with the continued fraction using
Eq.~(\ref{polstilCC}).  Again, we find good agreement of the results
for the dipole polarizability.

\subsection{The $^{22}$O nucleus}
\label{sec:22O} 

\begin{figure}[!ht]
\includegraphics[width=\linewidth,clip=]{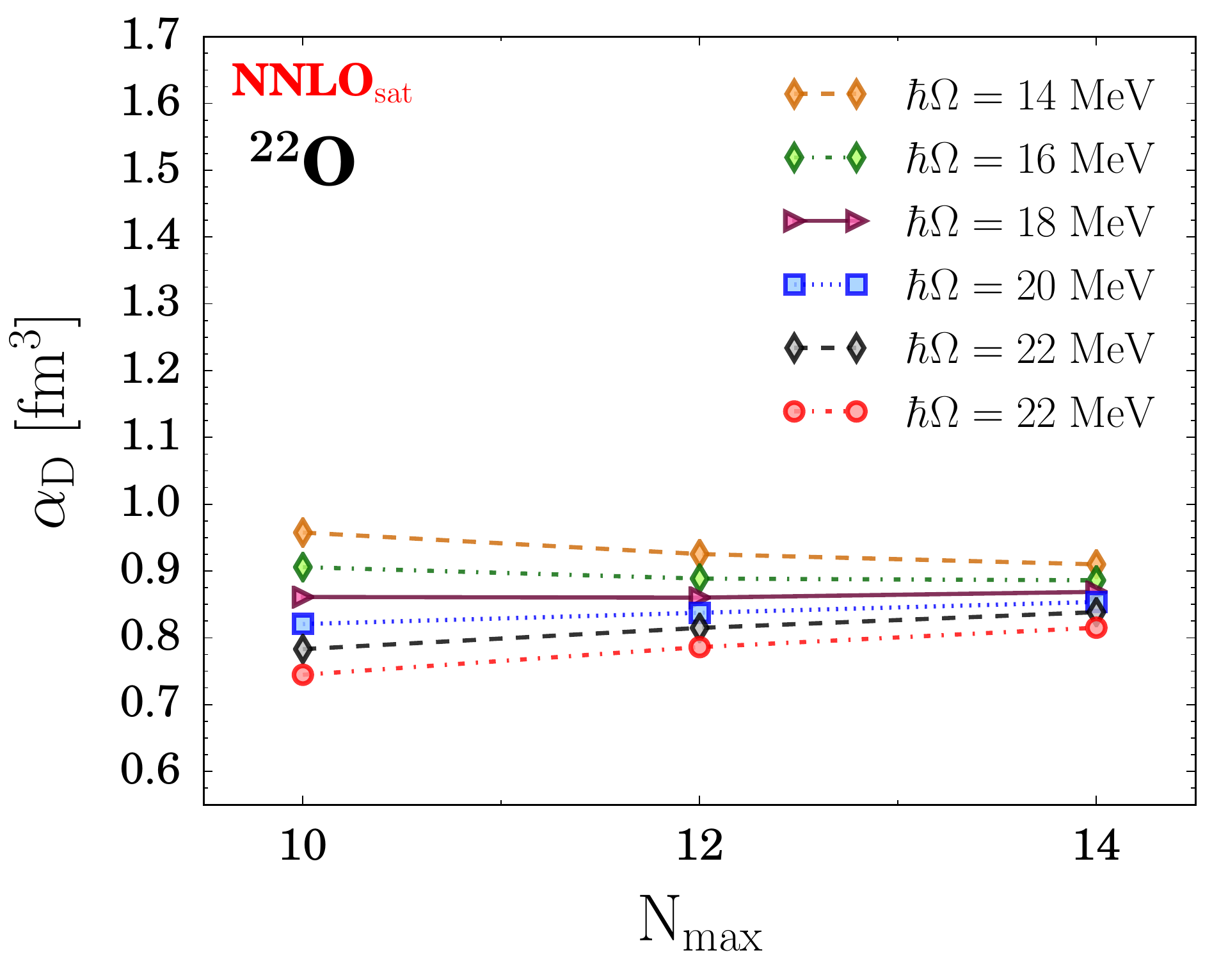}
\caption{(Color online) The electric dipole polarizability $\alpha_D$ in $^{22}$O as a function of the model space size $N_{\rm max}$. Different curves for different values of the underlying harmonic oscillator basis frequency $\hbar\Omega$ are shown. }
\label{fig:fig_pol_conv_22O}
\end{figure}

The dipole strength of the neutron-rich nucleus $^{22}$O was  measured by~\citeauthor{Leistenschneider2001}~\cite{ Leistenschneider2001} via Coulomb excitation in experiments at GSI. 
Figure~\ref{fig:fig_pol_conv_22O} shows the electric dipole polarizability as a function of the model space size of the
calculation. 
After having investigated various  frequencies, we find that $\hbar\Omega=18$ MeV is the best converging curve.  However, the convergence for different $\hbar\Omega$ is slower than what observed in lighter nuclei, resulting in a conservative uncertainty of about $8\%$ at $N_{\rm max}=14$. This might be because the excess neutrons in $^{22}$O are loosely bound, making the wave function more extended and thus the convergence slower.
We note that $\alpha_D$ of $^{22}$O is larger than for $^{16}$O. 

\begin{figure}[!ht]
\includegraphics[width=\linewidth,clip=]{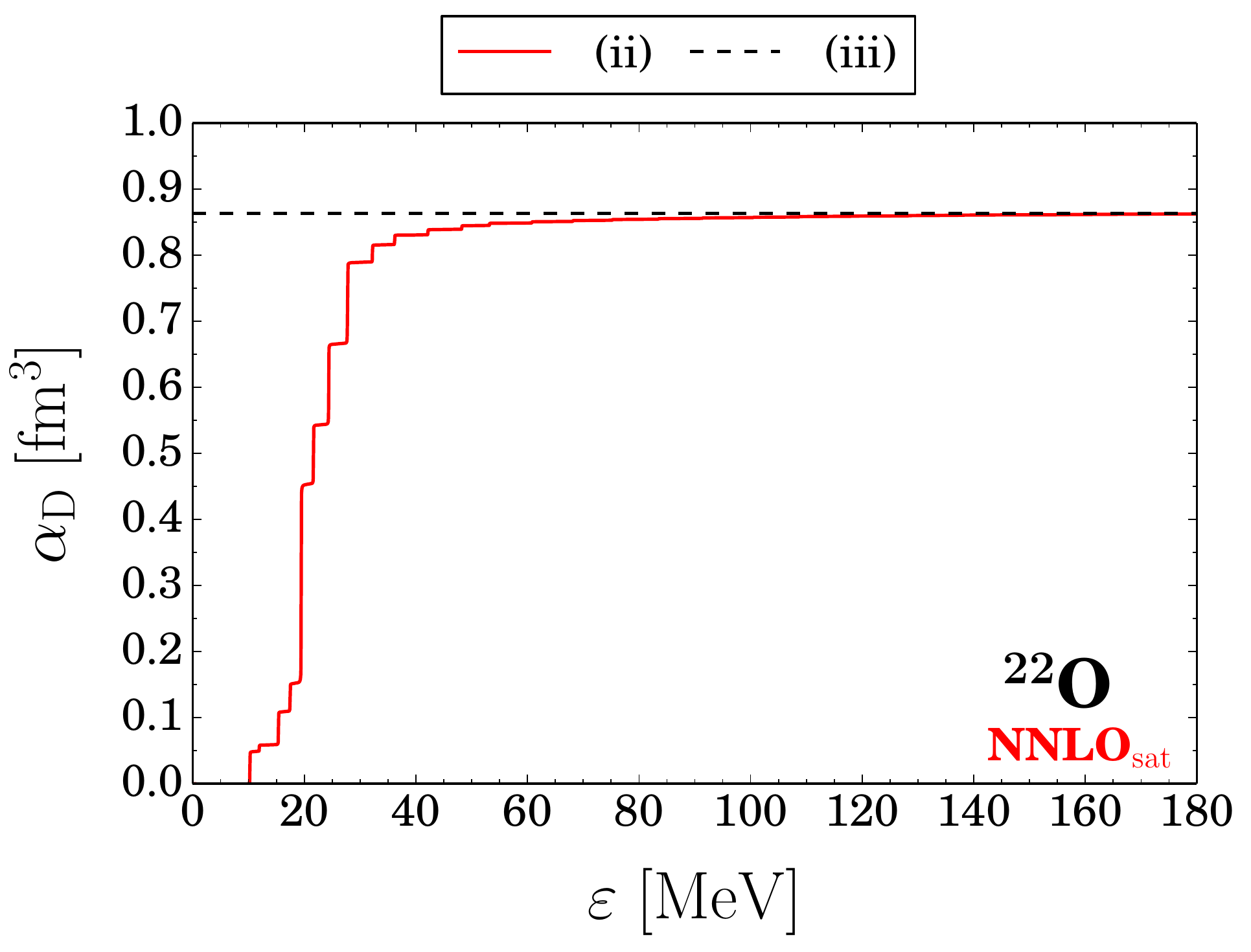}
\caption{(Color online) The electric dipole polarizability
  $\alpha_D(\varepsilon)$ in $^{22}$O as a function of the integration
  limit $\varepsilon$.  The red solid curve (ii) is calculated
  integrating the weighted LIT at small $\Gamma$ as in
  Eq.~(\ref{poldelta}); the black dashed line (iii) is obtained from
  the continued fraction of Eq.~(\ref{polstilCC}). Calculations are
  performed with $N_{\rm max}=14$ and $\hbar \Omega=18$ MeV.}
\label{fig:fig_polVSen_n2losat_22O}
\end{figure}
Finally, in Figure~\ref{fig:fig_polVSen_n2losat_22O} we show a
comparison between the methods (ii) and (iii) to calculate
$\alpha_D$. We used the largest model space and the fastest converging
frequency of $\hbar \Omega=18$~MeV and find good agreement between the
two methods.  Because the convergence of the LIT calculations is not
at sub-percent  level, we cannot presently obtain stable inversions and
include the method (i) in the comparison. Nevertheless, by looking at
the laddered curve we learn about the convergence of this sum rule as
a function of the energy. This will be used in the following
Subsection to make a comparison with the experimental data from
Ref.~\cite{Leistenschneider2001}.

\subsection{Comparison to experiment}
\label{subsec:correlations}

Table~\ref{compexp} compares theoretical results with experimental
data. We observe that for both $^4$He and $^{16}$O calculations are in
good agreement with the experimental data. For $^4$He the
experimental data is obtained by combining measurements from
Refs.~\cite{Arkatov74,Arkatov80,Pachucki07}. We also present a
comparison with other \textit{ab initio} results obtained with
hyperspherical harmonics~\cite{Gazit06,Ji13} and with the no-core
shell model~\cite{Stetcu09}. Because the experimental errorbar 
is quite large, all theoretical calculations are compatible with data.

\begin{table}[htb]
\caption{Theoretical values of $\alpha_D$ for different nuclei
  calculated with the NNLO$_{\rm sat}$ interaction in comparison to
  experimental data from~\cite{Arkatov74, Arkatov80,Pachucki07} and
  other calculations from Refs.~\cite{Stetcu09} (a), \cite{Gazit06}
  (b) and \cite{Ji13} (c) for $^4$He, to experimental data from
  Ref.~\cite{Ahrens75} for $^{16}$O. For $^{22}$O we compare to the
  value obtained integrating the data from
  Ref.~\cite{Leistenschneider2001} first over the whole energy range
  (d) and then only the first 3 MeV of the strength (e), corresponding
  to the low-lying dipole strength. Values are expressed in
  fm$^3$. The theoretical uncertainties of our calculations stem from
  the $\hbar\Omega$ dependence in the model space with $N_{\rm
    max}=14$.}
\label{compexp}
\begin{center}
\footnotesize
\renewcommand{\tabcolsep}{1.8mm}
\begin{tabular}{cll}
\hline\hline
{Nucleus}&{Theory}&{Exp}\\
\hline
{$^4$He}&{0.0735(1)}&{0.074(9)}\\
{ }&{0.0673(5)$^{a}$}&{}\\
{ }&{0.0655$^{b}$}&{}\\
{ }&{0.0651$^{c}$}&{}\\
{ }&{0.0694$^{c}$}&{}\\
\hline
{$^{16}$O}&{0.57(1)}&{0.585(9)}\\
\hline
{$^{22}$O}&{0.86(4)}&{0.43(4)$^{d}$}\\
{}&{0.05(1)}&{0.07(2)$^{e}$}\\
\hline \hline
\end{tabular}
\end{center}
\end{table}

For $^{16}$O the calculation of the dipole polarizability with the
NNLO$_{\rm{sat}}$ interaction overlaps with the experimental
value~\cite{Ahrens75}. This is an improvement compared to 
the previous calculation limited to NN interaction
only~\cite{Miorelli15}.

For the $^{22}$O nucleus, to compare our calculations with experimental data we integrate the experimental strength of Ref.~\cite{Leistenschneider2001} up to the available energy range of about 18~MeV above threshold, obtaining $\alpha_D^{exp} = 0.43(4)\ \rm{fm^3}$. This value is much lower than our calculated $\alpha_D^{th} = 0.86(4)\ \rm{fm^3}$ shown in Figure~\ref{fig:fig_polVSen_n2losat_22O}, which corresponds to the integration of the strength up to infinity. The theoretical result exceeds the experimental value by a factor of two and we also find that the integration of the theoretical strength over the first 18~MeV exhausts the 87\% of the polarizability sum rule.  On the other hand, \citeauthor{Leistenschneider2001} observed a PDR  extending for about 3~MeV above the neutron emission threshold of $S_n=6.85\ \rm{MeV}$. Integrating the data over this interval yields a dipole polarizability $\alpha_D^{\rm exp}(3~{\rm MeV}) = 0.07(2)\ \rm{fm^3}$. While our calculations in Figure~\ref{fig:fig_polVSen_n2losat_22O} does not  reproduce the experimental threshold, integration over the first 3 MeV of the strength and considering the different $\hbar\Omega$ frequencies yields $\alpha_D^{th}(\rm{PDR}) = 0.05(1)\ \rm{fm^3}$. This is consistent with the experimental result.\\

\begin{figure}[!ht]
\includegraphics[width=\linewidth,clip=]{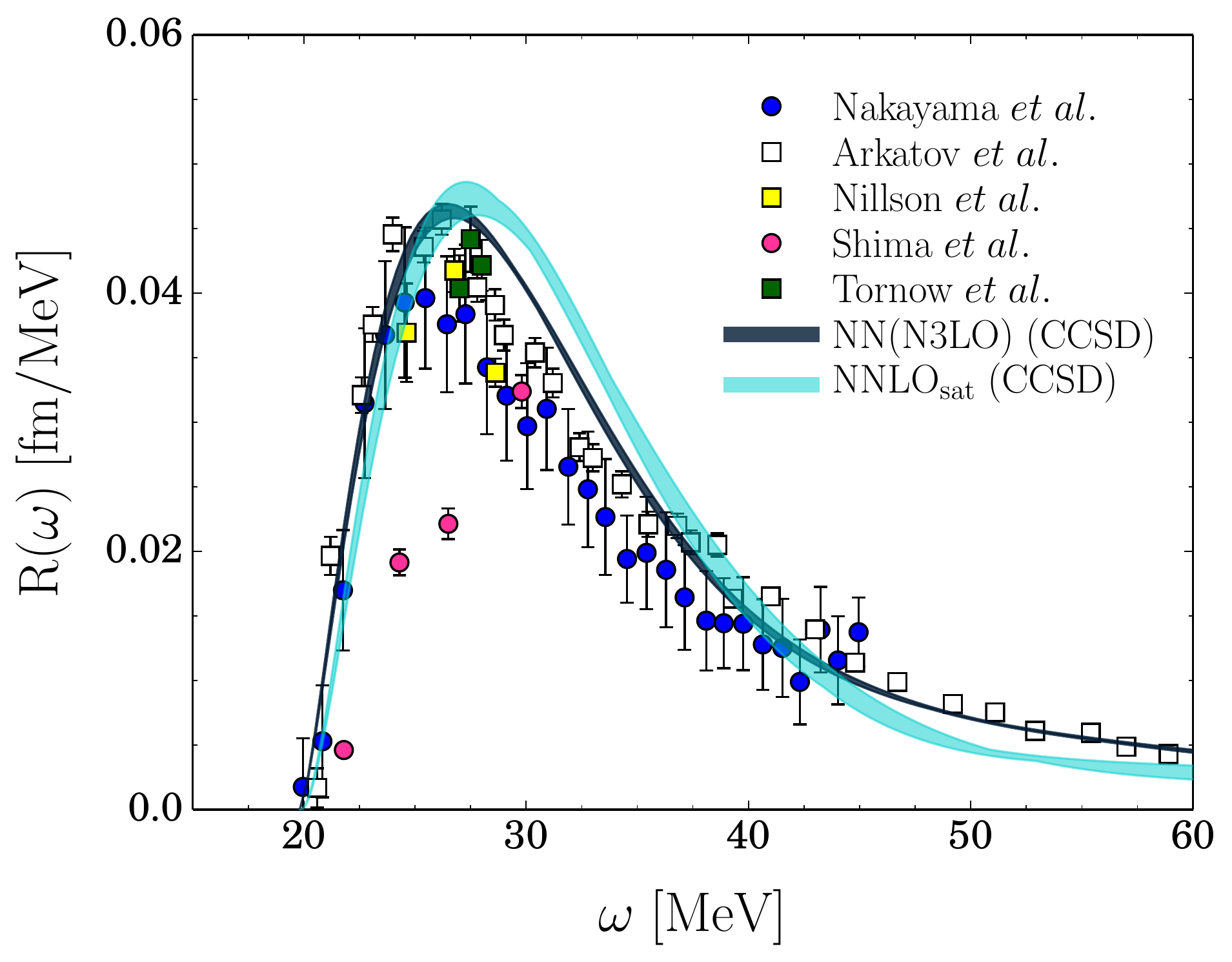}
\caption{(Color online) $^4$He photo-absorption response function calculated with different methods and interactions (see text for details) compared with experimental data from by~\citeauthor{nakayama}~\cite{nakayama} (blue circles), ~\citeauthor{Arkatov74}~\cite{Arkatov74,Arkatov80} (white squares), ~\citeauthor{Nilsson}~\cite{Nilsson} (yellow squares), ~\citeauthor{Shima2005}~\cite{Shima2005,Shima2010} (magenta circles) and ~\citeauthor{Tornow2012}~\cite{Tornow2012} (green squares). Theoretical curves are shifted on the experimental threshold.}
\label{fig:fig_resp_comp_4He}
\end{figure}

In Figure~\ref{fig:fig_resp_comp_4He} we show the response function of $^4$He. The response function is obtained from the inversion of the LIT as described in Refs.~\cite{Bacca13,Andreasi05,Efros99} and the width of the band is an estimate of the inversion uncertainty.
The dark band from Ref.~\cite{LITCC} is the result obtained with coupled cluster with singles-and-doubles (CCSD) using a NN interaction at next-to-next-to-next-to-leading order (N3LO)~\cite{Entem03}. The light  band represents the calculation of this work with NNLO$_{\rm sat}$~\cite{Ekstroem15} and it has been obtained by inverting the LIT with $\Gamma =10$ and 20~MeV calculated at $N_{\rm max}=14$ and $\hbar\Omega=22$~MeV. This is also the curve that has been integrated with method (i) in Figure~\ref{fig:fig_pol_vs_e_4He}. We find that the NNLO$_{\rm sat}$ response function, which includes three-nucleon forces, presents a larger peak with respect to other results with three-nucleon forces from Refs.~\cite{Gazit06PRL,Quaglioni2007}.
Finally, the theoretical results are compared with the experimental data by~\citeauthor{nakayama}~\cite{nakayama} (blue circles), ~\citeauthor{Arkatov74}~\cite{Arkatov74,Arkatov80} (white squares), ~\citeauthor{Nilsson}~\cite{Nilsson} (yellow squares), ~\citeauthor{Shima2005}~\cite{Shima2005,Shima2010} (magenta circles) and ~\citeauthor{Tornow2012}~\cite{Tornow2012} (green squares). 

\begin{figure}[!ht]
\includegraphics[width=\linewidth,clip=]{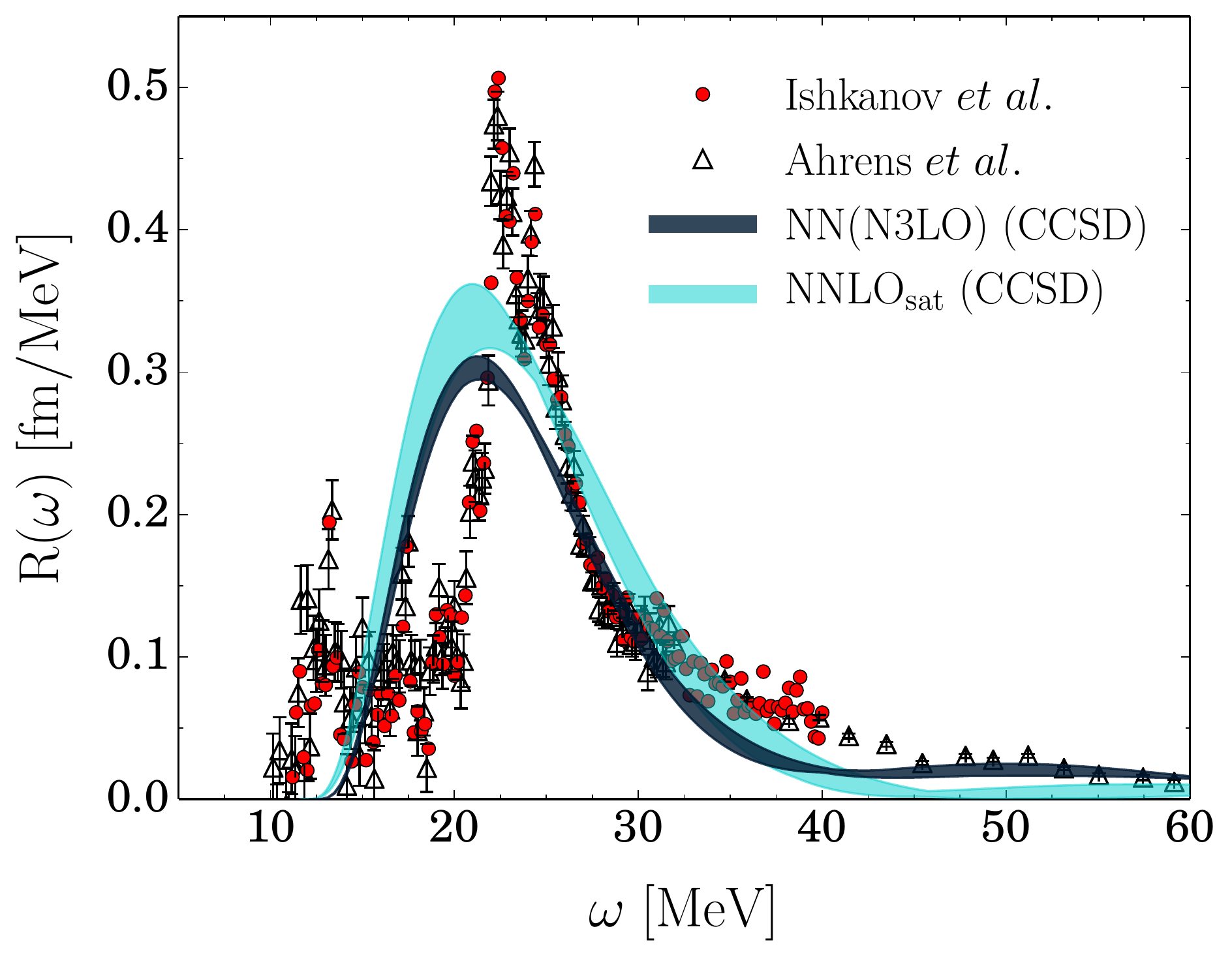}
\caption{(Color online) $^{16}$O photo-absorption response function calculated with coupled cluster with singles-and-doubles using a NN interaction only~\cite{Entem03,LITCC} (dark band) and NNLO$_{\rm sat}$~\cite{Ekstroem15} (light band). The red circles are the experimental data from~\citeauthor{Ishkhanov2002}~\cite{Ishkhanov2002} while the white triangles with error bars are the experimental results by~\citeauthor{Ahrens75}~\cite{Ahrens75}. Theoretical curves are shifted on the experimental threshold.}
\label{fig:fig_resp_comp_16O}
\end{figure}

In Figure~\ref{fig:fig_resp_comp_16O} we show the response function for $^{16}$O calculated with a NN interaction using CCSD~\cite{LITCC} (light band) and then with NNLO$_{\rm sat}$ (dark band). The calculations are compared with the experimental data from~\citeauthor{Ahrens75}~\cite{Ahrens75} (triangles with error bars) and~\citeauthor{Ishkhanov2002}~\cite{Ishkhanov2002} (red circles). The response function  with NNLO$_{\rm sat}$ has been obtained again by inverting the LIT with both $\Gamma=10$ and 20~MeV and at frequency $\hbar\Omega=22$~MeV. The large error band for the NNLO$_{\rm sat}$ results from the fact that the largest available model space size in our calculation, namely $N_{\rm max}=14$, is smaller than the  $N_{\rm max}=18$ used for the N3LO potential.  Nevertheless, it is interesting to see that three-nucleon forces enhance the strength, slightly improving the comparison with the experimental data.

Comparing Figure~\ref{fig:fig_pol_vs_e_4He} and~\ref{fig:fig_polVSen_n2losat_16O} with Figure~\ref{fig:fig_resp_comp_4He} and~\ref{fig:fig_resp_comp_16O} respectively, and taking into account the results summarized in Table~\ref{compexp}, it is clear that the polarizability is not very sensitive to the structure and shape of the response function, but rather to the distribution of the dipole strength at low energies.

\subsection{Correlations between $\alpha_D$ and $r_{ch}$}

Let us also attempt to probe systematic theoretical uncertainties that are due to the employed interaction by considering results from different families of Hamiltonians. Such an approach can help to correlate observables of interest, see Refs.~\cite{nogga2004,platter2005,Reinhard2010,Piekarewicz12,Roca-Maza2015,hagen2015,Calci16} for examples. To study such correlations, one needs a considerable number of different interactions, so that one can obtain results spanning a wide range of values for the observables under investigation. For this reason, we choose to use similarity renormalization group (SRG) ~\cite{bogner2007} and V$_{low-k}$ ~\cite{bogner2003} evolutions as a tool to generate a set of phase-shift equivalent two-body interactions. When adding three nucleon forces at next-to-next-to-leading order -- without considering the induced three-body forces --  the low-energy constants  were recalibrated on light nuclei observables~\cite{Hebeler11}. Finally, we also consider the newly developed NNLO$_{\rm sat}$ interaction~\cite{Ekstroem15}, which well reproduces radii~\cite{hagen2015}.
Various binding energies from NNLO$_{\rm sat}$ and other interactions of interest are shown in Refs.~\cite{Ekstroem15} and \cite{hagen2010b}.

We note that a correlation between the electric dipole polarizability and the nuclear charge radius $r_{ch}$ is expected from the nuclear droplet models~\cite{Myers77,Lipparini89} in heavy nuclei.  In what follows we investigate correlations between the dipole polarizability and charge radius in $^{16}$O and $^{40}$Ca using a variety of interactions. We base our calculations on NN forces and 3NFs from Refs.~\cite{Hebeler11,Ekstroem15}, and also consider computations limited to NN forces alone.

\begin{figure}[htp]
\subfloat{\includegraphics[clip,width=\columnwidth]{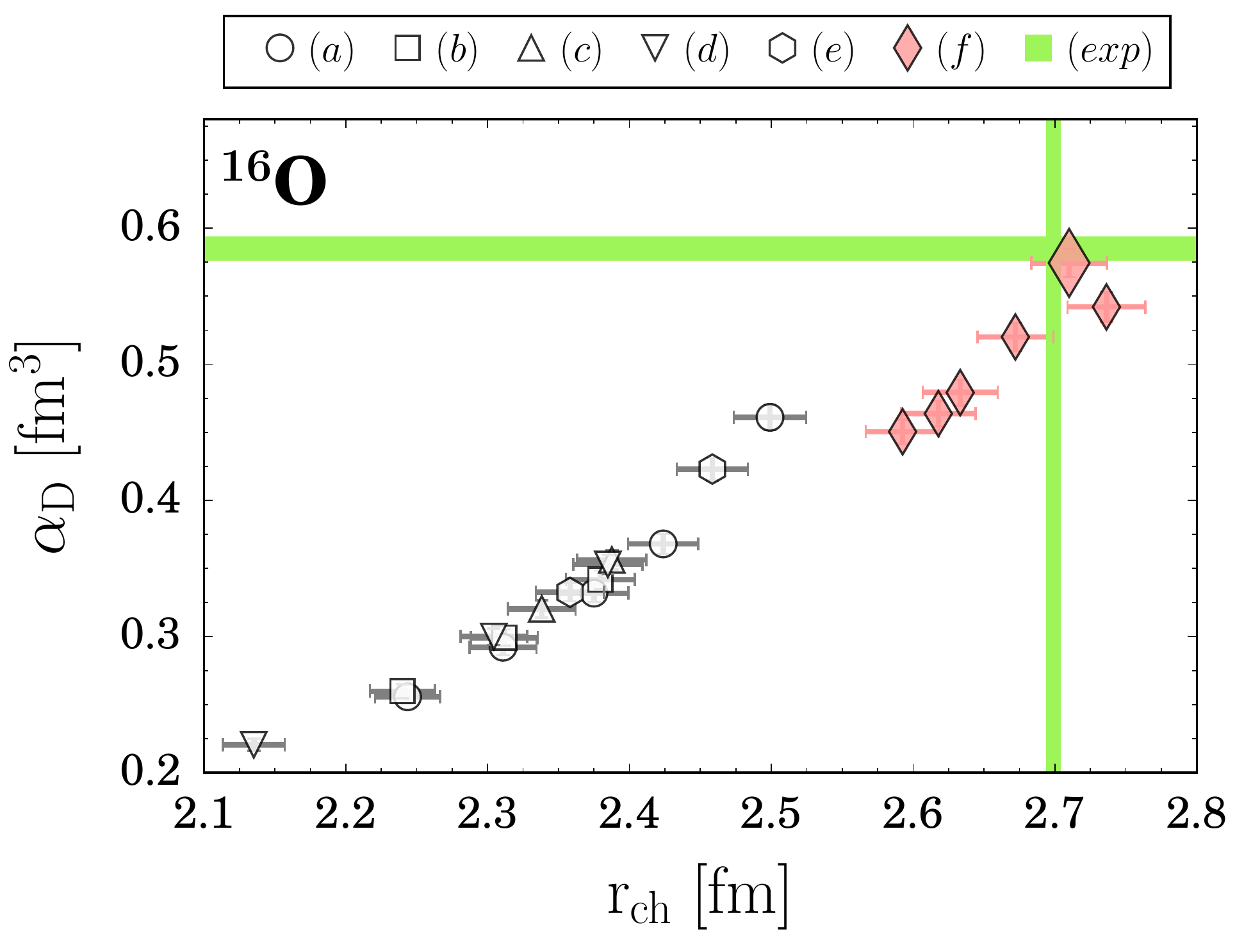}}

\subfloat{\includegraphics[clip,width=\columnwidth]{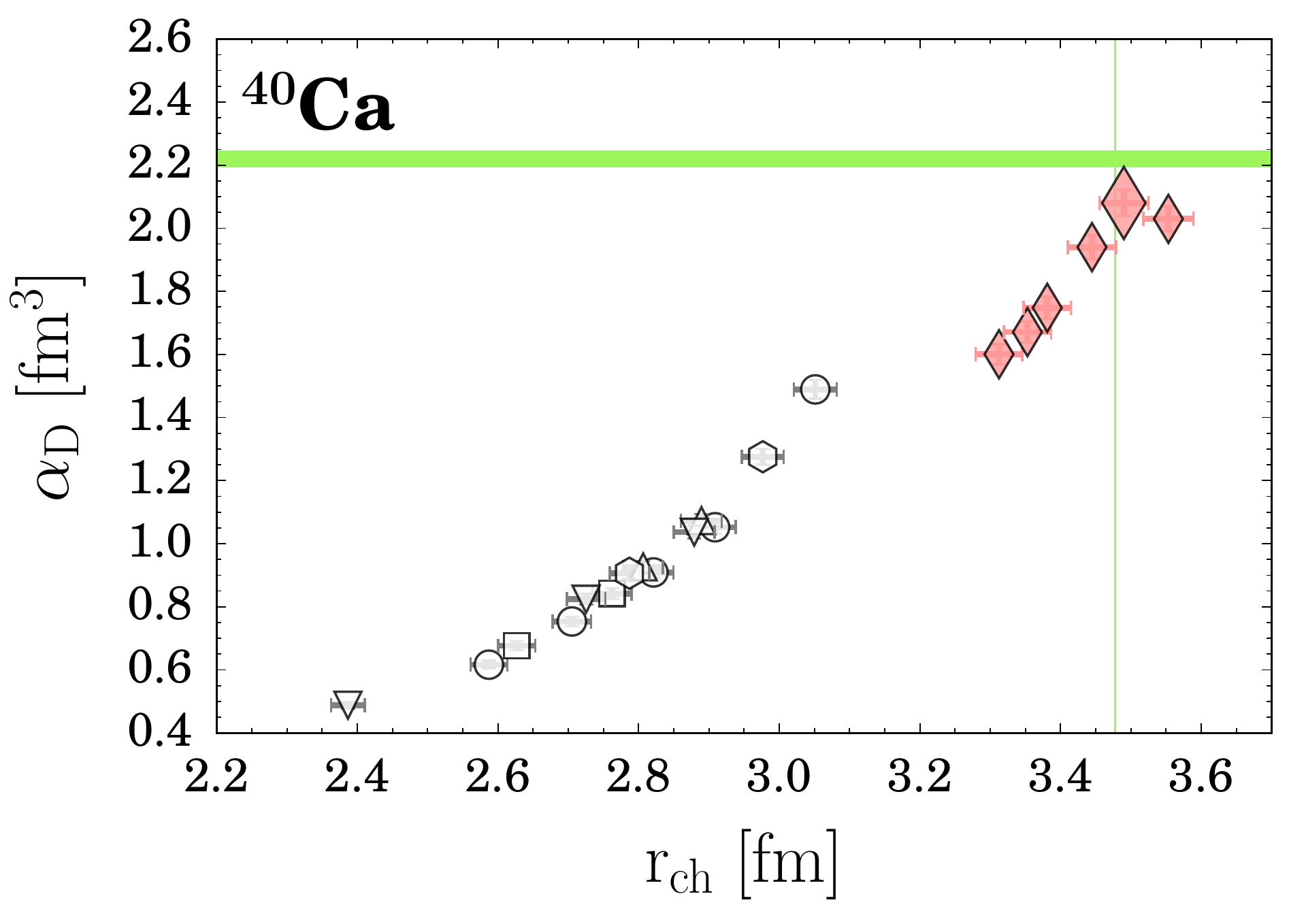}}

\caption{(Color online) $\alpha_D$ versus $r_{ch}$ in
  \textsuperscript{16}O and \textsuperscript{40}Ca.  Empty symbols
  refer to calculations with NN potentials only: $(a)$~SRG evolved Entem-Machleidt
  interaction~\cite{Entem03} with $\Lambda=500$~MeV/c and $\lambda = \infty,3.5,3.0,2.5$ and $2.0 \ \rm{fm^{-1}}$, $(b)$~SRG evolved Entem-Machleidt interaction~\cite{Entem03} with $\Lambda=600$~MeV/c and $\lambda = 3.5,3.0$ and $2.5\ \rm{fm^{-1}}$, $(c)$ SRG evolved CD-BONN~\cite{CDBONN} interaction with $\lambda = 4.0$ and $3.5\ \rm{fm^{-1}}$, $(d)$ V$_{low-k}$ evolved CD-BONN potentials with $\lambda = 3.0,2.5$ and $2.0\ \rm{fm^{-1}}$ and $(e)$ V$_{low-k}$-evolved AV18~\cite{AV18} interaction and $\lambda = 3.0$ and $2.5\ \rm{fm^{-1}}$. The red diamonds $(f)$ refer to calculations that include 3NF: the large one is from NNLO$_{\rm{sat}}$~\cite{Ekstroem15} and the others from chiral interactions as in Ref.~\cite{Hebeler11}. The green bands $(exp)$, show the experimental data~\cite{Ahrens75,Angeli13}.}
\label{fig:pol_vs_rch}
\end{figure}

Figure~\ref{fig:pol_vs_rch} shows $\alpha_D$  -- calculated with method (iii) -- as a function of $r_{ch}$ in $^{16}$O and $^{40}$Ca for various interactions.  The charge radii are based on the point-proton radii with contributions from nucleonic charge radii, see Ref.~\cite{hagen2015} for details.  Empty symbols correspond to calculations with NN potentials only. In particular, $(a)$ is obtained from SRG evolved Entem and Machleidt~\cite{Entem03} interaction with cutoff $\Lambda=500$~MeV and, in order of decreasing $r_{ch}$ values, $\lambda = \infty,3.5,3.0,2.5$ and $2.0\ \rm{fm^{-1}}$, while for $(b)$ we used the same interaction with cutoff $\Lambda=600$~MeV and, in order of decreasing $r_{ch}$ values, $\lambda = 3.5,3.0$ and $2.5\ \rm{fm^{-1}}$. The points $(c)$ represented with triangles pointing up are calculations with the SRG evolved CD-BONN~\cite{CDBONN} potential with, in order of decreasing $r_{ch}$ value, $\lambda = 4.0$ and $3.5\ \rm{fm^{-1}}$, while the triangles pointing down $(d)$ are calculations with the V$_{low-k}$~\cite{bogner2003} evolved CD-BONN interaction and $\lambda = 3.0,2.5$ and $2.0\ \rm{fm^{-1}}$. The hexagons $(e)$ are calculations with V$_{low-k}$-evolved AV18~\cite{AV18} interaction and $\lambda = 3.0$ and $2.5\ \rm{fm^{-1}}$, in order of decreasing radius.  The red diamonds $(f)$ are calculations including 3NFs. The larger red diamond is the value obtained with NNLO$_{\rm{sat}}$~\cite{Ekstroem15}, while the smaller ones are the potentials from Ref.~\cite{Hebeler11} also used for the calculations in $^{48}$Ca in Ref.~\cite{hagen2015}. 
The error bars for the calculations represent uncertainties arising both from the coupled-cluster truncation scheme and the model space truncations and are estimated to be of the order of 1\% for the charge radius and 2\% for the polarizability (see Ref.~\cite{hagen2015} for details). Finally, the green bands are the experimental values for the polarizability~\cite{Ahrens75} and the charge radius~\cite{Angeli13}, respectively.

We note that $\alpha_D$ and $r_{ch}$ are strongly correlated.  We also note that NN interactions alone systematically underestimate both $\alpha_D$ and $r_{ch}$ while the inclusion of 3NFs improves the agreement with data. The agreement with data is particularly good for the interaction NNLO$_{\rm{sat}}$. We note that one cannot blindly use a correlation between theoretical data points to extrapolate to experimental results. The data based on NN interactions, even when extrapolated with a simple linear or quadratic curve, does not meet the experimental values. In contrast, the results from NN and 3NFs can be interpolated (when e.g. the charge radius is known) to yield a sensible prediction for the dipole polarizability.

\section{Conclusions}
\label{sec:conclusions}

In conclusion, we employed integral transforms to compute the electric dipole polarizability in beta-stable nuclei and rare isotopes. This approach employs bound-state technology but takes the continuum properly into account.  We presented in detail the formalism for coupled-cluster calculations of $\alpha_D$ and computed the dipole polarizability in $^4$He, $^{16,22}$O, and $^{40}$Ca. Formulations as the dipole polarizability as an energy-weighted sum rule facilitate the comparison to data in cases where only lower-lying dipole strengths are measured.

The comparison with data reveals the important role of three-nucleon forces, and results based on the NNLO$_{\rm{sat}}$ interaction agree well with data in $^{4}$He and $^{16}$O, and $^{40}$Ca. For the neutron-rich $^{22}$O, the low-lying dipole strength within 3~MeV of threshold theoretical results are consistent with data, while the total theoretical dipole strength is about a factor of two larger than what can be computed from the available data.  Further investigation is needed to study the shape of the low-energy strength distribution.

Finally, we studied $^{16}$O and $^{40}$Ca with different two- and three-body interactions and observed a strong correlation between the dipole polarizability and the charge radius.  Such a correlation could be useful to predict either of the two observables, when only one of them is experimentally known. Work in this direction is underway for heavier nuclei, such as $^{68}$Ni and $^{90}$Zr.

\begin{acknowledgments}
TRIUMF receives federal funding via a contribution agreement with the National Research Council of Canada. This work was supported in parts by the Natural Sciences and Engineering Research Council (Grant number SAPIN-2015-00031), the US-Israel Binational Science Foundation (Grant
No.~2012212), the Pazy Foundation, the MIUR grant PRIN-2009TWL3MX, the
Office of Nuclear Physics, U.S.~Department of Energy under Grants
Nos.~DE-FG02-96ER40963 (University of Tennessee) and DE-SC0008499
(NUCLEI SciDAC collaboration), and the Field Work Proposal ERKBP57 at
Oak Ridge National Laboratory. Computer time was provided by the
Innovative and Novel Computational Impact on Theory and Experiment
(INCITE) program. This research used resources of the Oak Ridge
Leadership Computing Facility located in the Oak Ridge National
Laboratory, supported by the Office of Science of the U.S.~Department
of Energy under Contract No.  DE-AC05-00OR22725, and computational
resources of the National Center for Computational Sciences, the
National Institute for Computational Sciences, and TRIUMF.
\end{acknowledgments}

\bibliography{refs}

\end{document}